\documentclass[twocolumn, secnumarabic, floatfix, nofootinbib, nobalancelastpage, longbibliography]{revtex4-2}

\def\TitleOfPaper{Generalized Gradient Approximation Made Thermal}
\def\preprintlinklocation{http://dft.uci.edu + PAPER REF}

\usepackage{amsmath}
\usepackage{physics}
\usepackage{amssymb}
\usepackage{float}
\usepackage[dvipsnames]{xcolor}

\definecolor{TITLECOL}{rgb}{0.05,0.25,0.85}
\definecolor{CONTENTSCOL}{rgb}{0.1,0.2,0.7}
\definecolor{URLCOL}{rgb}{0,0.52,0.83}
\definecolor{LINKCOL}{rgb}{0.05,0.5,0}
\definecolor{CITECOL}{rgb}{0.25,0,0.48}
\definecolor{SECOL}{rgb}{0.07,0.31,0.80}
\definecolor{SSECOL}{rgb}{0.26,0.19,0.75}

\newcommand{\coloredtitle}[1]{\title{\textcolor{TITLECOL}{#1}}}
\newcommand{\coloredauthor}[1]{\author{\textcolor{CITECOL}{#1}}} 
\renewcommand{\sec}[1]{\section{\textcolor{SECOL}{#1}}}
\newcommand{\ssec}[1]{\subsection{\textcolor{SSECOL}{#1}}}

\usepackage{graphicx}

\def\PublicationsLink{http://dft.uci.edu/publications.php}
\def\preprintlink{ \href{\preprintlinklocation}{\TitleOfPaper} }
\def\preprinttext{~}

\usepackage{fancyhdr}
\makeatletter
\fancypagestyle{titlepage}
{	\lhead{\textsc{\preprinttext\ ~ \href{\PublicationsLink}{~}}}
	\chead{}
	\rhead{\preprintlink}
	\lfoot{}}
\makeatother
\def\preprintlink{ 
	\href{\preprintlinklocation}
        {
~}
	}

\definecolor{Green}{rgb}{0.016,0.627,0}
\definecolor{Plum}{rgb}{0.17,0,0.45}
\definecolor{LBlue}{rgb}{0,0.34,0.45}
\definecolor{Sepia}{rgb}{0.37,0.17,0.02}
\definecolor{BurntOrange}{rgb}{0.78,0.39,0}

\usepackage{hyperref}
\hypersetup{
    breaklinks=true,
    bookmarksopen=true,
    bookmarksnumbered=true,
    colorlinks=true,
    linkcolor=LINKCOL,
    linktocpage=true,
    citecolor=CITECOL,
    filecolor=magenta,      
    urlcolor=URLCOL,
}

\def\bea{\begin{eqnarray}}
\def\eea{\end{eqnarray}}
\def\ben{\begin{equation}}
\def\een{\end{equation}}
\def\benu{\begin{enumerate}}
\def\enu{\end{enumerate}}

\def\bei{\begin{itemize}}
\def\eei{\end{itemize}}
\def\beit{\begin{itemize}}
\def\eit{\end{itemize}}
\def\benu{\begin{enumerate}}
\def\enu{\end{enumerate}}

\def\n{n}

\def\sss{\scriptscriptstyle\rm}

\def\g{_\gamma}

\def\1var{(\bx_1...\bx\N)}

\def\br{{\bf r}}

\def\bx{{x}}

\def\x{_{\sss X}}
\def\c{_{\sss C}}
\def\s{_{\sss S}}
\def\xc{_{\sss XC}}

\def\N{_{\sss N}}

\def\LDA{^{\rm LDA}}

\def\PBE{^{\rm PBE}}
\def\ltPBE{^{\rm ltPBE}}

\def\unif{^{\rm unif}}

\def\unpol{^{\rm unpol}}
\def\pol{^{\rm pol}}

\def\up{_\uparrow}
\def\dn{_\downarrow}

\def\sph_int{ {\int d^3 r}}

\begin{document}
\sf
\coloredtitle{\TitleOfPaper}

\coloredauthor{John Kozlowski}
\affiliation{Department of Chemistry, University of California, Irvine, CA 92697, USA}

\coloredauthor{Dennis Perchak}
\affiliation{Department of Chemistry, University of California, Irvine, CA 92697, USA}

\coloredauthor{Kieron Burke}
\affiliation{Department of Chemistry, University of California, Irvine, CA 92697, USA}
\affiliation{Department of Physics and Astronomy, University of California, Irvine, CA 92697, USA}

\date{\today}


\begin{abstract}
Using the methodology of conditional-probability density functional theory, and several mild assumptions, we calculate the temperature-dependence of the Perdew–Burke–Ernzerhof (PBE) generalized gradient approximation (GGA).  This numerically-defined thermal GGA reduces to the local approximation in the uniform limit and PBE at zero temperature, and can be fit reasonably accurately (within 8\%) assuming the temperature-dependent enhancement is independent of the gradient. This locally thermal PBE satisfies both the coordinate-scaled correlation inequality and the concavity condition, which we prove for finite temperatures. The temperature dependence differs markedly from existing thermal GGA's.
\end{abstract}

\maketitle


\textbf{Introduction:} Density functional theory (DFT) is utilized throughout modern science, wherever electronic structure is important, and has enormous impact in materials simulations \cite{JSP16, GVKH18, PEE20, TB21} and quantum chemistry \cite{T14, JFFT11, XWL23}. Almost all such calculations employ the Kohn-Sham (KS) scheme \cite{KS65}, in which only the exchange-correlation (XC) energy need be approximated as a functional of electronic (spin) densities \cite{DG90}. The quality of results generated depends crucially on that approximation, and many hundreds of such approximations are readily available in modern codes \cite{LSOM18}.

Warm dense matter (WDM) includes nuclear fusion at the national ignition facility, interiors of gas giant planets, and matter under extreme shock conditions at the Sandia Z-machine or SLAC’s free electron laser \cite{PHDPD18, NRB08, NHKFRB08, WKACRC20}. Simulations of WDM have been greatly improved by the use of KS-DFT at finite temperature \cite{HRD08, KRDM08, RMCHM10, MDC22, Dea23, HBN23}. Mermin showed that an equilibrium grand canonical density and free energy can be found this way, but the unknown XC free energy is now also a function of temperature \cite{M65}. Most successful simulations performed today, however, use existing ground-state XC approximations. While their success can be understood, the missing thermal dependence is an uncontrolled error that may significantly impact their results.

\begin{figure}[t]
    \centering
    \includegraphics[width=0.95\linewidth]{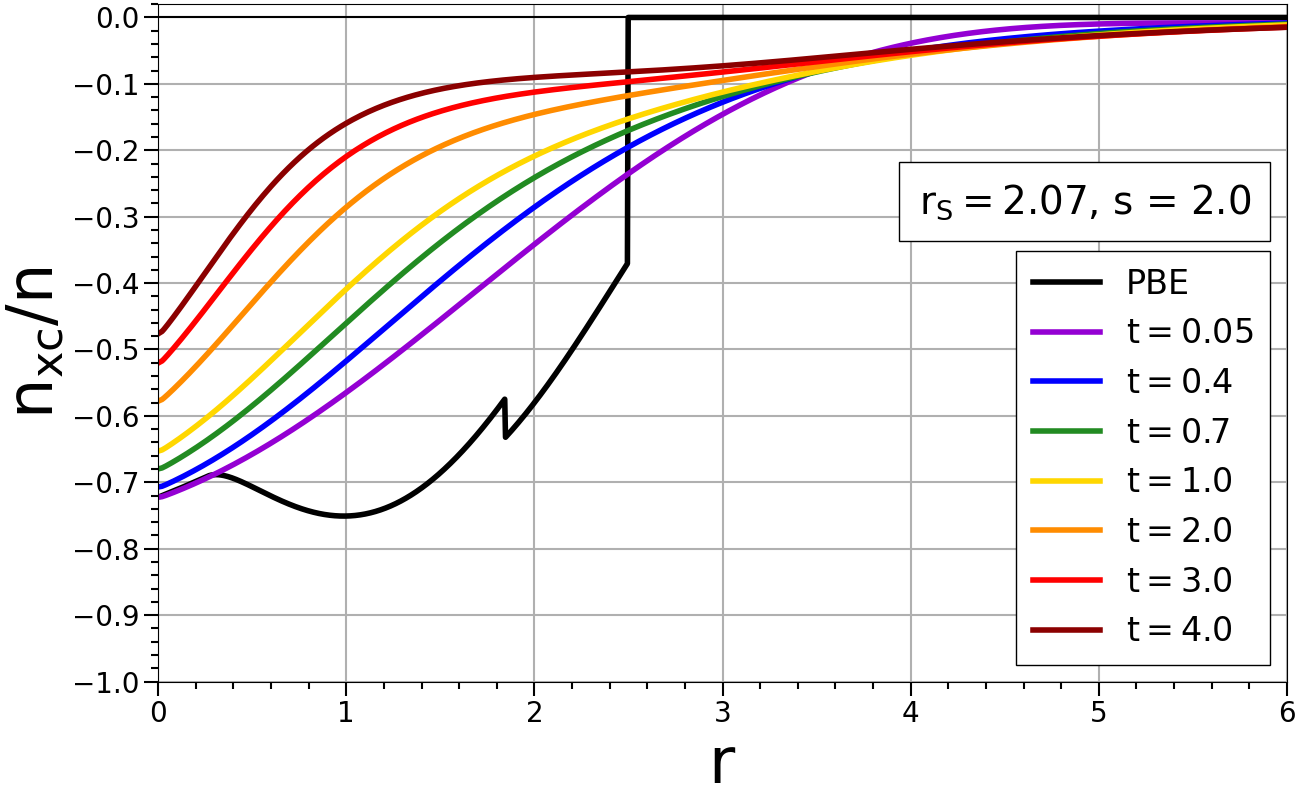}
    \vspace{-6pt}
    \caption{Typical temperature-dependent XC hole densities. Black denotes the ground-state real-space cutoff GGA hole density, while the CP-DFT XC hole at different reduced temperatures is depicted in various colors. As $t\rightarrow0$, CP-DFT yields an XC hole density with the same on-top value and energy as PBE. See text for definitions.}
    \vspace{-12pt}
    \label{fig:XCHoleAlu}
\end{figure}

\begin{figure*}
    \centering
    \includegraphics[width=0.8925\linewidth]{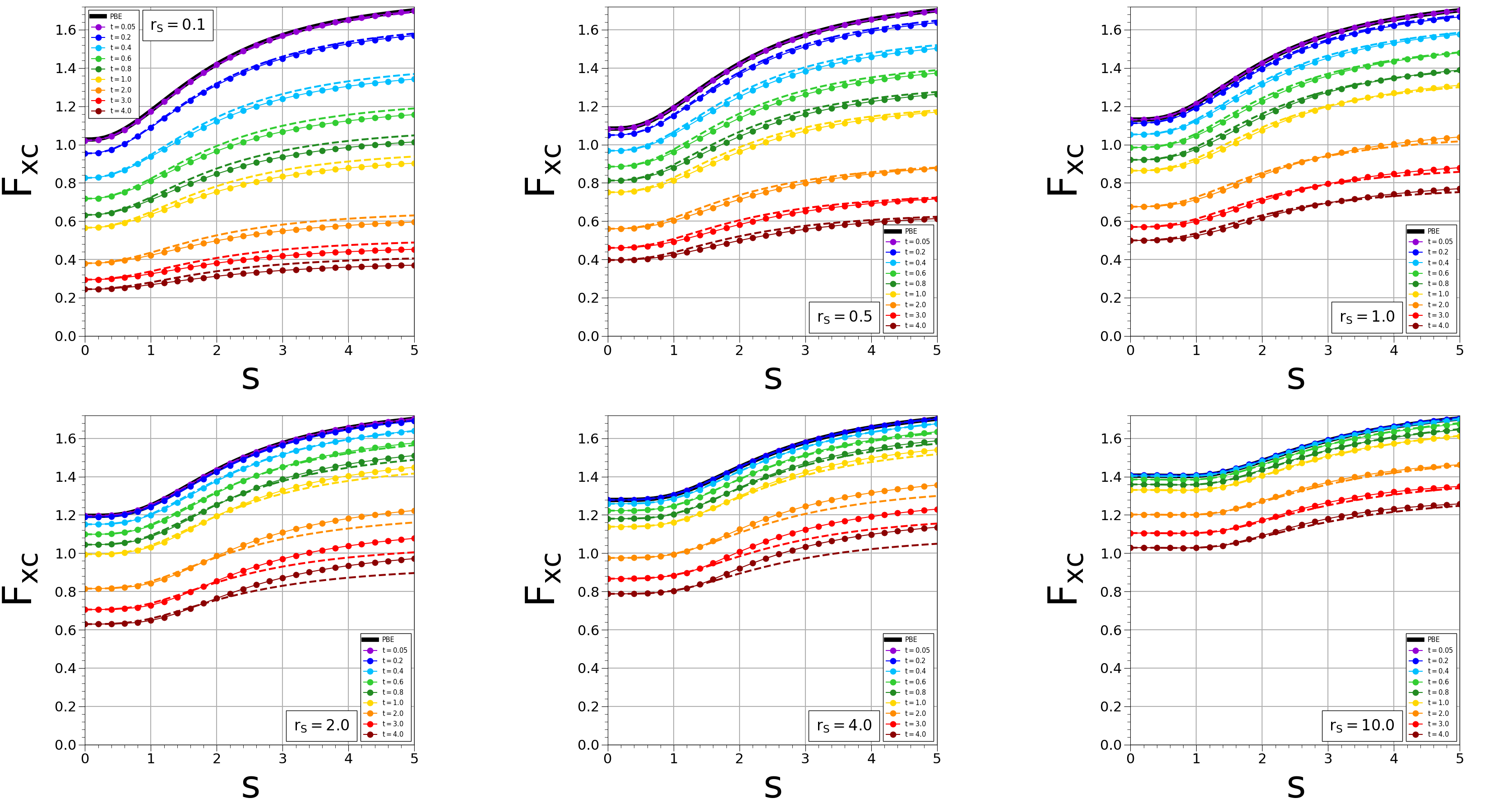}
    \vspace{-6pt}
    \caption{Temperature-dependent XC enhancement factors plotted as functions of the dimensionless gradient $s$ for unpolarized systems of various $r\s$ values. Here we denote the results of CP-DFT calculations at different temperatures using data points of various colors, while the ltPBE approximation (Eq.~\ref{ltPBE}) is represented by dashed curves. The ground-state PBE XC enhancement factor is shown in black, which the CP-DFT results approach as $t \rightarrow 0$. More plots (including those for fully polarized systems) are shown in Figs.~S1 and S2 of the supplemental material.}
    \vspace{-12pt}
    \label{fig:ltPBEhalf_Unp}
\end{figure*}

At the local density approximation (LDA) level \cite{KS65}, there is a long history of parameterizing the XC free energy of a uniform electron gas \cite{KSDT14, GDSMFB17, KDT18, KTD19} as a function of temperature. But modern materials simulations require at least a generalized gradient approximation (GGA) level of accuracy to achieve chemical specificity, which is why the Perdew–Burke–Ernzerhof (PBE) approximation is used in most such simulations. The underlying rationale for the PBE functional is a detailed model of the XC hole, and the exact conditions used in its derivation were chosen to (approximately) recover the numerical results of imposing exact conditions on the hole, not the energy directly.

The present work gives the result of calculating the XC hole of PBE as a function of temperature, using the recently-invented conditional-probability (CP) density functional theory \cite{PCWB22, MPPEQWB20, PMB22} (see Refs.~\citenum{GP01, GS05, GS07} for its precursors). In this procedure one first finds, at zero temperature, the CP potential that generates the PBE hole at a given density and gradient. Then, by ignoring the temperature dependence of the CP potential, one can calculate the XC hole using finite-temperature KS-DFT. Throughout this work, the Wigner-Seitz radius $r\s = (3 / 4 \pi n)^{1/3}$, the dimensionless density gradient $s = |\nabla n| / 2 k_F n$ with Fermi wavevector $k_F = (3 \pi^2 n)^{1/3}$, and the reduced temperature $t = T / T_F$ with Fermi temperature $T_F = k_F^2 / 2 k_B$. Fig.~\ref{fig:XCHoleAlu} illustrates our work, showing the XC hole of aluminum ($r\s = 2.07$) with a nonzero gradient at various temperatures. This result not only demonstrates the behavior of PBE's XC hole at WDM conditions, but also graphically depicts the diminishing effect temperature has on the XC free energy, a quantity directly related to the XC hole via integration. By construction, the resulting (numerically-defined) free energy GGA reduces to PBE at zero temperature. Moreover, very little error is introduced by ignoring temperature dependence in the CP potential, as has already been shown for the uniform gas \cite{MPPEQWB20}.

An important second step is to create a simple parameterization of the resulting functional, so that it can be easily implemented and utilized. We find that the simplest possibly physical realistic case (referred to throughout this work as {\em locally thermal PBE})
\begin{equation}
    F\xc\ltPBE (r\s, s, t) = \frac{F\xc\unif (r\s, t)}{F\xc\unif (r\s)} \times F\xc\PBE (r\s, s),
    \label{ltPBE}
\end{equation}
matches the output from our CP-DFT calculation to within a few percent, examples of which are shown in Fig.~\ref{fig:ltPBEhalf_Unp} for unpolarized systems of various $r\s$ values. Such a simple parameterization automatically satisfies several key exact conditions, and is trivial to implement in existing codes. Thus, the effects of these new thermal corrections can be immediately tested in any WDM simulation currently making use of PBE.

A prior attempt to incorporate finite temperature-dependence makes use of exact conditions \cite{KDT18}, but yields a strikingly different temperature dependence.

\textbf{Background:} CP-DFT makes use of an approximate potential (referred to as the CP potential \cite{PCWB22}) at each point in space to calculate the conditional probability density, which can be integrated to yield the XC energy via
\begin{equation}
    E\xc = \frac{1}{2} \int_0^1 d\lambda \int d^3r \int d^3 r' \, \frac{n(\br)\left[\tilde{n}_{\br}^\lambda(\br') - n(\br)\right]}{|\br - \br'|},
    \label{Exc}
\end{equation}
where $\tilde{n}_{\br}^\lambda(\br')$ is the conditional probability density (the probability of finding an electron at $\br'$, given already having found one at $\br$), and $n\xc(\br, \br') = \int_0^1 d\lambda \left[\tilde{n}_{\br}^\lambda(\br') - n(\br)\right]$ is the coupling-constant averaged XC hole density. Thus, the XC hole of a system may be written as the difference between the CP density and standard electronic density, each of which are determined self-consistently using independent KS-DFT calculations. The reliability of this approach is readily explained through this feature - the extraction of XC free energies via Eq.~\ref{Exc} eliminates functional error from our energies, as no approximate XC density functional need be evaluated \cite{VSKSB19}. The quality of our result is directly attributed to the accuracy of the two self-consistent densities, which have previously been shown to be highly accurate (even when approximate functionals are utilized) \cite{KSB13}. In this work, we use this methodology to generate a thermal GGA.

The CP potential for an unpolarized uniform electron gas of $N - 1$ electrons may be written as \cite{MPPEQWB20}
\begin{equation}
    v\s(\br) = \Delta\tilde{v}(r) + \int d^3 r' \, \frac{\tilde{n}(r') - n_0}{|\br - \br'|} + v\xc\LDA[\tilde{n}](r),
    \label{CPpotential1}
\end{equation}
where $n_0 = N/V$ is a constant, and
\begin{equation}
    \Delta\tilde{v}(r) = \frac{1}{2r}\left[1 + \erf{\left(\frac{r}{r\s}\right)}\right] + A(r\s, s) e^{-r^2 / 2\sigma(r\s, s)^2}
    \label{CPpotential2}
\end{equation}
approximates the effect of removing one electron from the system. The first term of this approximation~\cite{MPPEQWB20, GP01} classically approximates the effect of the missing electron, while the second term works to recreate the correct high-density limit for uniform ($s=0$) systems. 

In our procedure, we modify this Gaussian term to approximate the CP potential of non-uniform systems, using it as a means of recreating the characteristic qualities of the PBE XC hole. The amplitude and width of a given Gaussian is numerically fit for each $r\s$ and $s$ value, such that the result of a CP-DFT calculation yields an XC hole with the same (1) on-top value and (2) XC energy as PBE's hole as the reduced temperature $t$ approaches zero. For fitting purposes, we generate PBE's XC hole using the damped, numerical procedure in Refs. \citenum{PBW96} \& \citenum{BPW98}. This makes use of the second-order gradient expansion of the XC hole, but enforces numerical cut-offs to ensure satisfaction of the exchange/correlation sum rules. In principle, the exact CP potential is temperature-dependent, but we expect this temperature dependence to have little effect, vanishing in the high temperature limit. In the uniform limit \cite{MPPEQWB20}, this approximation yields errors of order $5\%$.

There exists no equivalent CP potential approximation (Eq.~\ref{CPpotential2}) for polarized systems, and the derivation of one is beyond the scope of this work. In its place, we generate data for fully polarized systems by making use of the exchange hole's spin-scaling relation $n\x\pol[n](\br,\br') = n\x\unpol[2n](\br,\br')$. Thus, we are able to recover the polarized exchange hole exactly through scaling, but miss the correlation hole, which must be approximated with an added corrective potential such that Eq.~\ref{CPpotential2} becomes:
\begin{equation}
    \Delta\tilde{v}(r) = \frac{1}{2r}\left[1 + \erf{\left(\frac{r}{r\s}\right)}\right] + A_1 e^{-r^2 / 2\sigma_1^2} + A_2 e^{-r^{3/4} / 2\sigma_2^2}
    \label{CPpotential3}
\end{equation}
We note that $A_1$ and $\sigma_1$ are found first, in the same manner as before (but this time using a scaled density). The last term of Eq.~\ref{CPpotential3} is then introduced (via a second numerical search to find $A_2$ and $\sigma_2$) so that the characteristics of the polarized version of PBE's XC hole are recreated. All numerical parameters ($A_1$, $A_2$, $\sigma_1$, $\sigma_2$) are functions of both $r\s$ and $s$, and given as .txt files in the supplemental material. Note the Fermi temperature of a fully polarized system $T_F\pol = 2^{2/3} T_F\unpol$.

Fig.~\ref{fig:ltPBEhalf_Unp} depicts CP-DFT results for unpolarized systems of $r\s = 0.1,\, 0.5,\, 1,\, 2,\, 4,\, 10$ and $0.05 \leq t \leq 4$; equivalent calculations have been performed for fully polarized systems of $r\s = 0.1,\, 0.5,\, 1,\, 2,\, 4$. The resulting XC hole densities and XC enhancement factors for all CP-DFT calculations are plotted in the first section of the supplemental material found at https://doi.org/XX.XXXX/XXXXXXXXXXX, with the latter being also given as .txt files.

The output of the CP-DFT procedure, applied to the numerically-defined cut-off GGA hole, yields a well-defined thermal GGA, which we refer to as CPTGGA (conditional probability thermal GGA). The results depend weakly on the specific choices made here. We anticipate both refining those choices and carefully parameterizing the output in the future. For the present, the locally thermal PBE defined by Eq.~\ref{ltPBE} captures most of the temperature-dependence in CPTGGA. Moreover, this simple prescription can be applied to any approximate ground-state functional, as a suggested temperature dependence, which could then be tested.

\begin{figure}[t]
    \centering
    \includegraphics[width=0.95\linewidth]{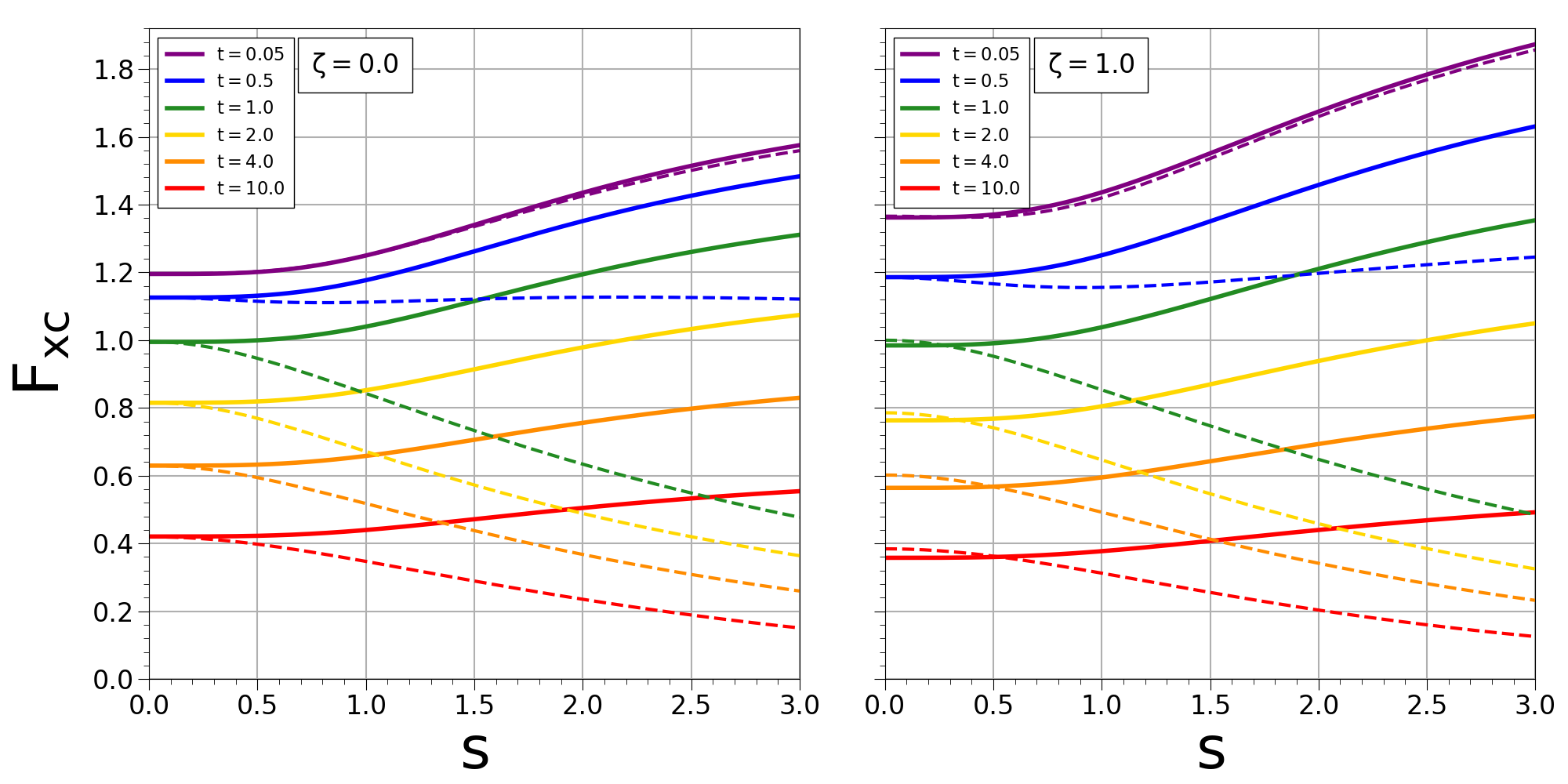}
    \vspace{-10pt}
    \caption{Temperature-dependent XC enhancement factors plotted as functions of the dimensionless gradient $s$. Here we set $r\s = 2$, showing the results for both unpolarized (left) and fully polarized (right) systems. Solid curves represent $F\xc\ltPBE$, while dashed curves represent $F\xc^{\rm KDT16}$.}
    \vspace{-12pt}
    \label{fig:Fxc}
\end{figure}

\textbf{Analysis:} We begin by emphasizing again that any conclusions drawn from this work are based on a temperature-dependent model for PBE's XC hole, which is shown in Fig.~\ref{fig:XCHoleAlu} for an unpolarized system with $r\s = 2.07$ and $s = 2$. As $t$ is increased, it is clear that the negative on-top value of the hole (and its surrounding density) monotonically increases, resulting in smaller XC free energies. Although the hole shrinks with temperature, the sum rule $\int n\xc (\br, \br + \br') \, d^3r' = -1$ is satisfied by the CP-DFT XC hole for all temperatures by construction.

When discussing GGA XC free energies, it is useful to define the temperature-dependent XC enhancement factor $F\xc$ as a multiple of the ground-state uniform electron gas exchange energy:
\begin{equation}
    A\xc(r\s, \zeta, s, t) = \int d^3r \, \epsilon\x\unif(n) \, F\xc (r\s, \zeta, s, t)
\end{equation}
Here $A\xc$ is the XC free energy, $\zeta = (n\up - n\dn) / n$ is the relative spin polarization, and $\epsilon\x\unif(n) = -3k_F/4\pi$. This allows us to not only perform direct comparisons to the ground-state PBE approximation, but also to the thermal GGA previously proposed by Karasiev {\em et al.} (referred to here as KDT16) \cite{KDT18}. Throughout this work, we implement ltPBE via Eq.~\ref{ltPBE} using the thermal LDA parameterization proposed by Groth {\em et al.} \cite{GDSMFB17}, which is given for all $\zeta$.

In Fig.~\ref{fig:Fxc} we plot XC enhancement factors as functions of the dimensionless gradient $s$, comparing ltPBE to KDT16 at various temperatures. First, note that both GGA's converge to PBE as $t\rightarrow0$, with negligible differences for $t = 0.05$; in the low temperature limit, both GGA's predict $F\xc$ to monotonically increase with respect to the gradient. Similarly, both GGA's approach nearly the same $F\xc$ value as $s\rightarrow0$, as both reduce to their corresponding thermal LDA (which have noticeable differences for $\zeta=1$) in this limit. The two approximations however predict quite different behaviors for nonzero temperature/gradient values. Although KDT16 recreates the monotonically increasing behavior of PBE at $t=0$, this trend flips and becomes monotonically decreasing with density gradient for warm/hot temperatures. In contrast, ltPBE by definition is a simple temperature-dependent multiple of the ground-state $F\xc\PBE$ curve, and thus inherits its curvature. As the temperature is increased further, the importance of the gradient on the XC free energy becomes less prominent, so both approximations vary less with $s$. Thus, there are significant differences between the approximations in the warm dense matter regime, particularly for systems with large gradients.

\begin{figure}[t]
    \centering
    \includegraphics[width=0.98\linewidth]{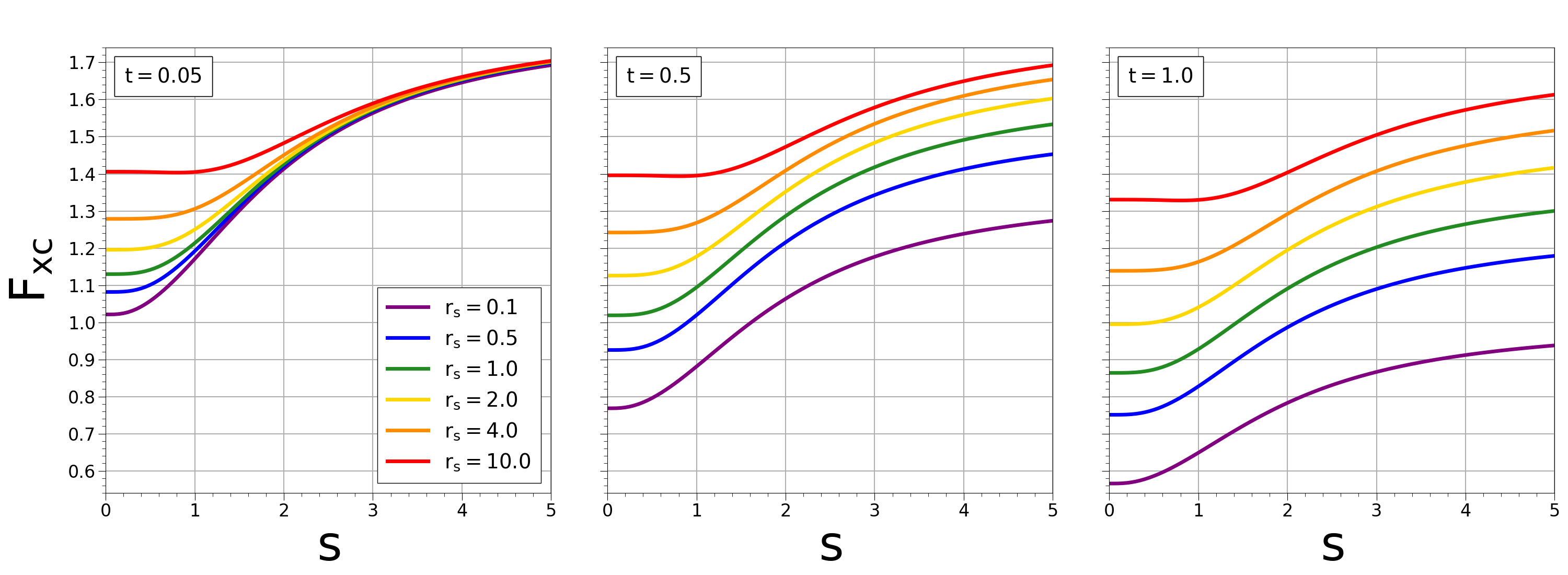}
    \vspace{-11pt}
    \caption{Several ltPBE XC enhancement factors plotted as functions of the dimensionless gradient $s$ for unpolarized systems of various $r\s$ values, with each plot having a different fixed reduced temperature $t$. The thermal coordinate scaling inequality (Eq.~\ref{FxcScaling}) mandates these curves not cross for any $t$, which the ltPBE approximation is shown to satisfy.}
    \vspace{-12pt}
    \label{fig:CSIltPBE}
\end{figure}

We now turn our attention to two exact conditions, both of which were omitted in the formulation of the ``Strongly Constrained and Appropriately Normed” (SCAN) semilocal functional \cite{SRP15, PB23}, and are readily generalizable to finite temperatures. Using uniform coordinate scaling inequalities at nonzero temperature, the XC free energy has been shown to scale according to the relation \cite{PPFSBBG11}
\begin{equation}
    A\xc^T [n_\gamma] \geq \gamma A\xc^{T / \gamma^2} [n], \quad\quad (\gamma \geq 1)
    \label{AxcScaling}
\end{equation}
where $\gamma$ represents the coordinate scaling parameter, and the scaled density $n_\gamma (\br) = \gamma^3 n (\gamma \br)$. This results in an exact condition for thermal generalized gradient approximations:
\begin{equation}
    F\xc (r\s, \, s, \, t) \geq F\xc (r\s', \, s, \, t) \quad\quad (r\s' \leq r\s)
    \label{FxcScaling}
\end{equation}
Note that the reduced temperature remains constant here, since $t = T / T_F = 2k_BT/(3\pi^2n)^{2/3}$ is scaled equally on both sides of Eq.~\ref{AxcScaling}. This signifies that the XC enhancement factors of different densities should not cross for any chosen reduced temperature. In Fig.~\ref{fig:CSIltPBE} we plot $F\xc$ curves at fixed values of $t$ for the ltPBE approximation, illustrating its satisfaction of this condition. The ratio $F\xc\unif (r\s, t) / F\xc\unif (r\s)$ decreases with $r\s$, by inspection. This fact, combined with PBE's satisfaction of Eq.~\ref{FxcScaling} at zero temperature~\cite{PB23}, guarantees ltPBE satisfies it by construction for all temperatures. Equivalent plots and analysis of KDT16, which is shown to violate this condition, are given in Fig.~S5 of the supplemental material.

\begin{figure}[t]
    \centering
    \includegraphics[width=0.95\linewidth]{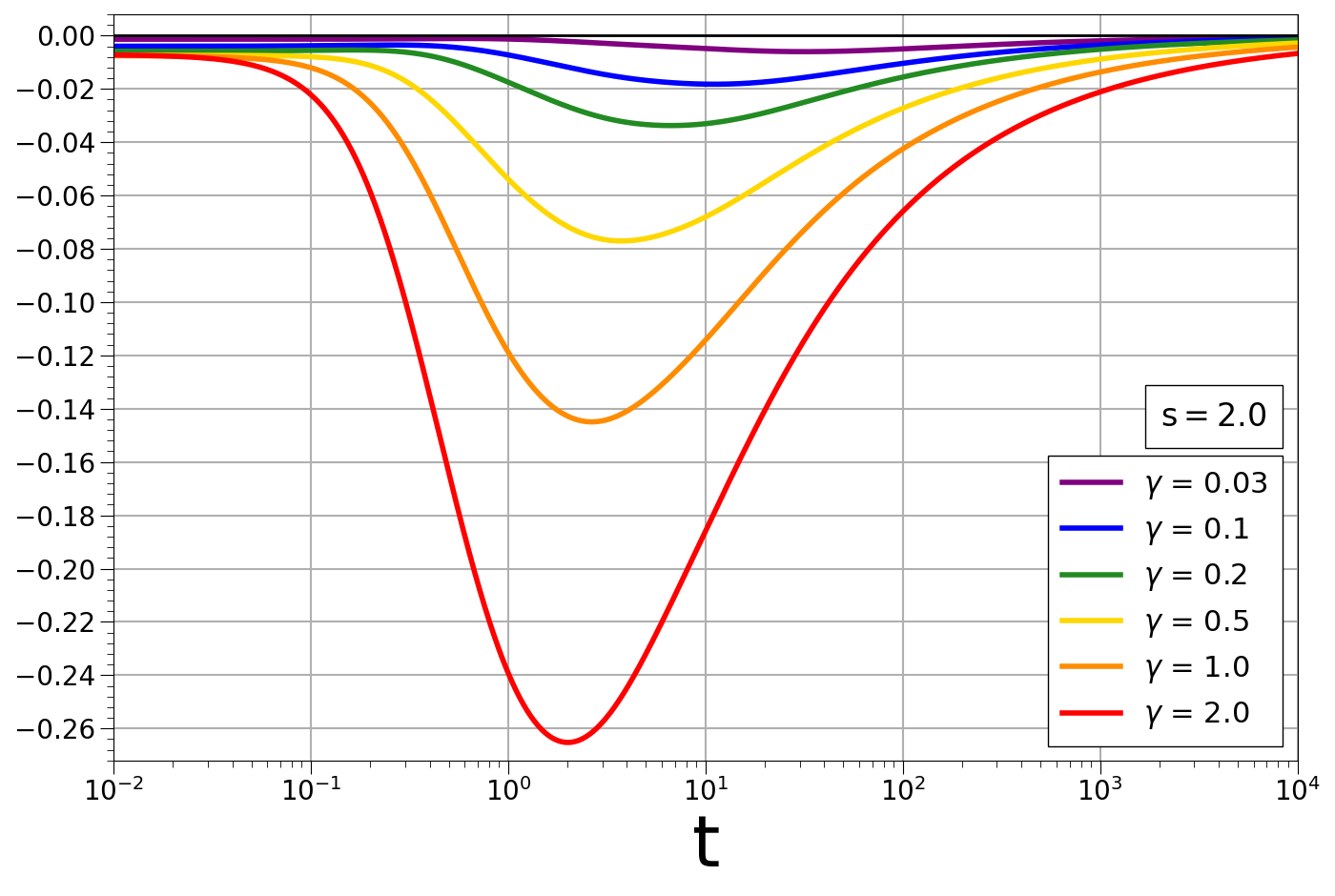}
    \vspace{-15pt}
    \caption{The thermal concavity condition inequality (Eq.~\ref{CC}) plotted for ltPBE with initial density $n=1$ and various $\gamma$. All curves are negative, indicating the exact condition is satisfied.}
    \vspace{-12pt}
    \label{fig:CChalf}
\end{figure}

The second exact condition we test involves the second derivative of the correlation free energy (the concavity condition) with respect to the coordinate scaling parameter $\gamma$. First derived as an exact condition of the ground state correlation energy~\cite{LP93}, it has recently been extended to ensemble DFT by Scott {\em et al.}~\cite{SKCPB23}. Here we show the extension of this condition to the correlation free energy of thermal ensembles, where states are occupied using temperature-dependent Fermi weights. To achieve this, we make note of the coordinate scaling relations \cite{PPFSBBG11}
\begin{equation}
    K\c^T[n_\gamma] \leq \gamma^2 K\c^{T / \gamma^2}[n], \quad A\c^T[n_\gamma] \geq \gamma A\c^{T / \gamma^2}[n], \quad (\gamma \geq 1)
    \label{UCSI}
\end{equation}
where $A\c^T[n]$ is the correlation free energy, and $K\c^T[n] = T\c^T[n] - T S\c^T[n]$ is its kentropic contribution. Considering $\gamma=1+\epsilon$ and taking $\epsilon \rightarrow 0$, we find differential versions of Eq.~\ref{UCSI} to be
\begin{equation}
    \frac{d}{d \gamma} \left\{\frac{K\c^T[n_\gamma]}{\gamma^2}\right\} \leq 0, \quad \frac{d}{d \gamma} \left\{\frac{A\c^T[n_\gamma]}{\gamma}\right\} \geq 0.
\end{equation}
Taking these relations in conjunction with the thermal virial theorem (Eq.~23 of Ref.~\citenum{PB16}), we find the concavity constraint on the correlation free energy
\begin{equation}
    \bigg( 2 - 2\gamma\frac{d}{d \gamma} + \gamma^{2}\frac{d^{2}}{d \gamma^{2}} \bigg) \, A\c^T[\n\g] \leq 0.
    \label{CC}
\end{equation}
In Fig.~\ref{fig:CChalf}, we plot an example of Eq.~\ref{CC} for ltPBE with initial density $n=1$ and $s=2$. All curves, regardless of the value of $\gamma$, are negative and thus satisfy Eq.~\ref{CC}. Contour plots showing more examples for both ltPBE and KDT16 are given in Fig.~S6 of the supplemental material. While ltPBE is shown to satisy Eq.~\ref{CC} by inspection, KDT16 is found to only satisfy the condition as $s\rightarrow0$.

While both exact conditions are found to be satisfied by ltPBE, we have checked the first only by inspection for CPTGGA. Noting that the ltPBE free exchange energy is found using a modified version of Eq.~\ref{ltPBE}
\begin{equation}
    F\x\ltPBE (s, t) = F\x\unif (t) \times F\x\PBE (s),
    \label{ltPBEx}
\end{equation}
where by definition the ground-state $F\x\unif = 1$, we find that ltPBE satisfies all exact conditions built into KDT16, other than the weakly inhomogeneous electron gas limit~\cite{KDT18}. Since PBE violates the gradient expansion, it follows that CPTGGA (and therefore ltPBE) does as well.

\textbf{Conclusions:} In this work we have generated the temperature-dependence of the PBE ground-state functional through a series of CP-DFT calculations capturing the characteristics of the PBE XC hole at zero temperature. We have provided a simple ansatz (Eq.~\ref{ltPBE}) that allows for our temperature-dependent GGA to be readily implemented in standard DFT codes. Analysis of our ltPBE approximation reveals striking differences with previously proposed thermal GGA's. We have shown that this simple approximation satisfies crucial exact constraints of the XC free energy, namely the temperature-dependent XC coordinate scaling inequality (Eq.~\ref{FxcScaling}) and concavity condition (Eq.~\ref{CC}). WDM simulations are ongoing to determine the importance of ltPBE thermal gradient effects, as there may be significant improvements from incorporating the temperature-dependence of the XC energy explicitly. Furthermore, locally thermalized versions of other ground-state functionals may offer significant improvements to the accuracy of WDM simulations, and should be tested in future work.

\textbf{Acknowledgements:} J.K. and K.B. acknowledge support from NSF award number CHE-2154371. We thank Attila Cangi and Tobias Dornheim for helpful discussions.

\bibliographystyle{apsrev4-2}
\bibliography{Master}

\label{page:end}
\end{document}


\sf
\coloredtitle{\TitleOfPaper}

\coloredauthor{John Kozlowski}
\affiliation{Department of Chemistry, University of California, Irvine, CA 92697, USA}

\coloredauthor{Dennis Perchak}
\affiliation{Department of Chemistry, University of California, Irvine, CA 92697, USA}

\coloredauthor{Kieron Burke}
\affiliation{Department of Chemistry, University of California, Irvine, CA 92697, USA}
\affiliation{Department of Physics and Astronomy, University of California, Irvine, CA 92697, USA}

\date{\today}


\begin{abstract}
Below we provide supplemental figures accompanying the main text, including various plots depicting the accuracy of the locally-thermal Perdew–Burke–Ernzerhof (ltPBE) ansatz relative to conditional-probability density functional theory (CP-DFT) data, and its satisfaction of two exact conditions. When useful, we make comparisons to other thermal generalized gradient approximations in the literature, namely the one proposed by Karasiev {\em et al.} (referred to here as KDT16).
\end{abstract}

\maketitle

\tableofcontents


\sec{CP-DFT Results}

\ssec{XC Enhancement Factors}

All CP-DFT XC enhancement factor data has been included as .txt files for systems of various densities, gradients, and temperatures. We include files for unpolarized systems: $r\s = 0.1,\, 0.5,\, 1,\, 2,\, 4,\, 10$ for $0 \leq s \leq 5$ and  $0.05 \leq t \leq 4$; and for fully polarized systems: $r\s = 0.1,\, 0.5,\, 1,\, 2,\, 4$ for $0 \leq s \leq 3$ and  $0.05 \leq t \leq 4$.

\begin{figure}[t]
    \centering
    \includegraphics[width=\linewidth]{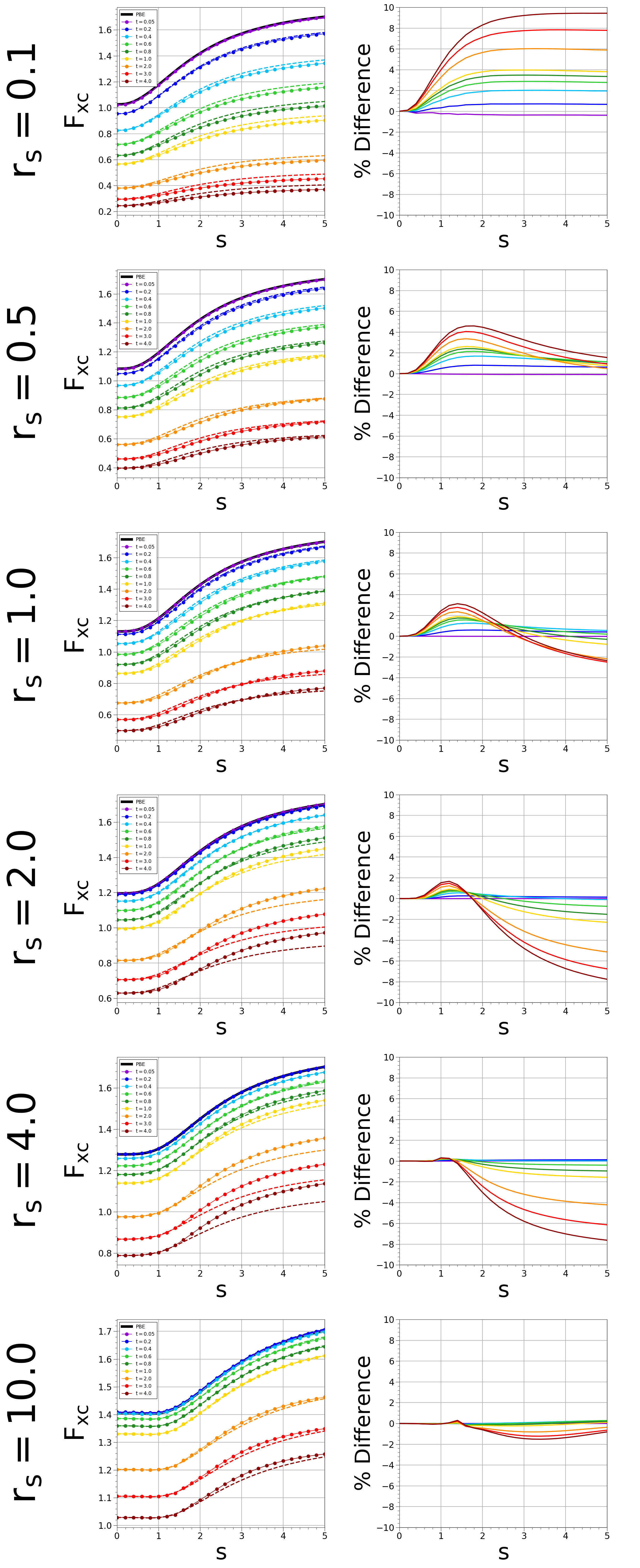}
    \caption{Grid of plots depicting temperature-dependent XC enhancement factors as functions of the dimensionless gradient $s$ for unpolarized systems of various $r\s$ values. The data points represent CP-DFT calculations connected by solid lines, while the ltPBE approximation (Eq.~1) is represented by the dashed curves. The corresponding percent deviation of the ltPBE ansatz is included, largely found to be $\leq 8\%$ (but often much less) in the tested temperature range.}
    \label{fig:ltPBEfull_Unp}
\end{figure}

\begin{figure}[t]
    \centering
    \includegraphics[width=\linewidth]{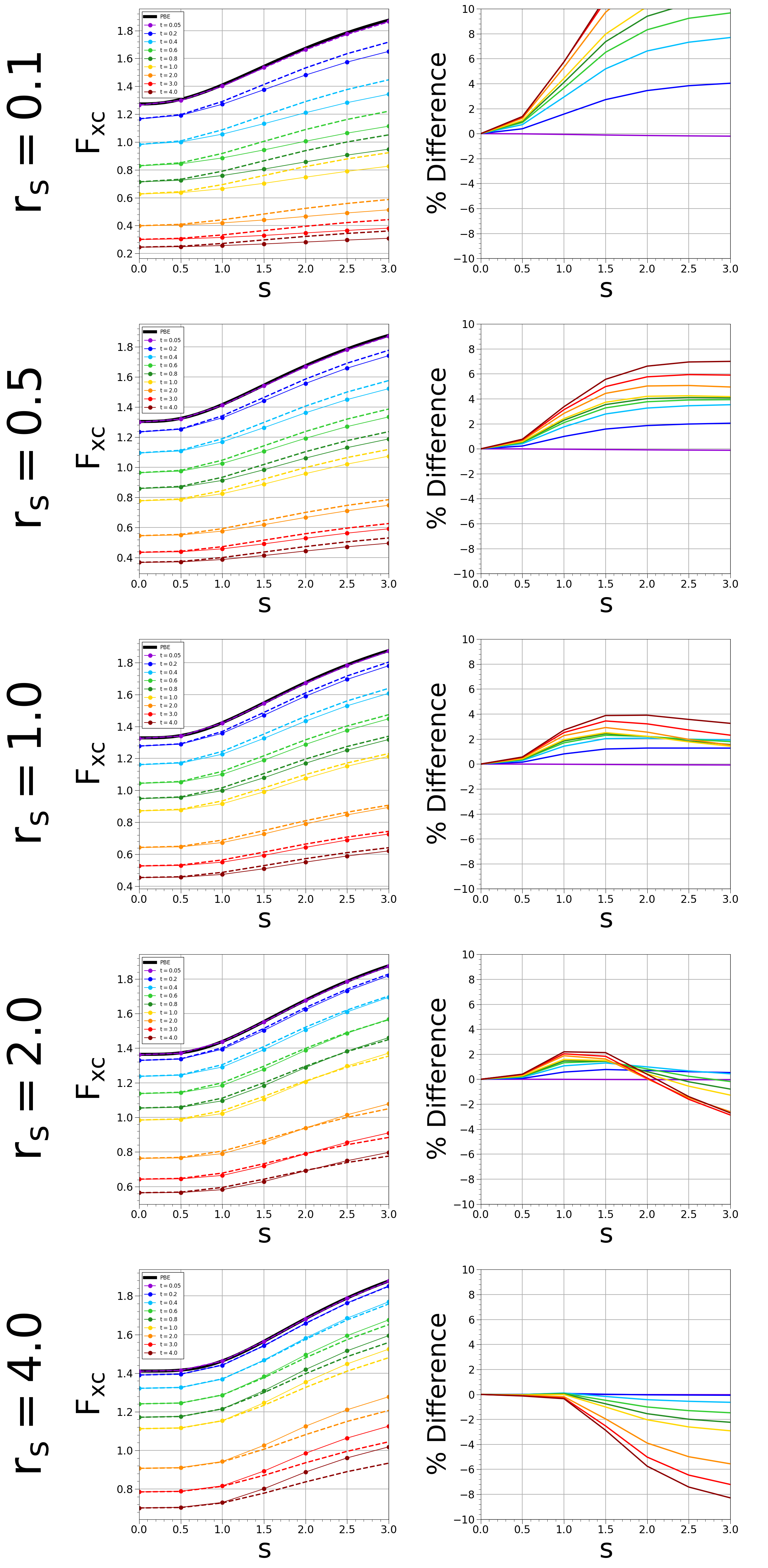}
    \caption{Grid of plots depicting temperature-dependent XC enhancement factors as functions of the dimensionless gradient $s$ for fully polarized systems of various $r\s$ values. The data points represent CP-DFT calculations connected by solid lines, while the ltPBE approximation (Eq.~1) is represented by the dashed curves. The corresponding percent deviation of the ltPBE ansatz is included, largely found to be $\leq 8\%$ (but often much less) in the tested temperature range.}
    \label{fig:ltPBEfull_Pol}
\end{figure}

The ltPBE approximation is depicted alongside CP-DFT data in the following figures, showing its performance for each density at select reduced temperatures. Fig.~\ref{fig:ltPBEfull_Unp} depicts the results for unpolarized data, while Fig.~\ref{fig:ltPBEfull_Pol} depicts the results for fully polarized data.

\ssec{XC Hole Densities}

\begin{figure*}
    \centering
    \includegraphics[width=0.84\linewidth]{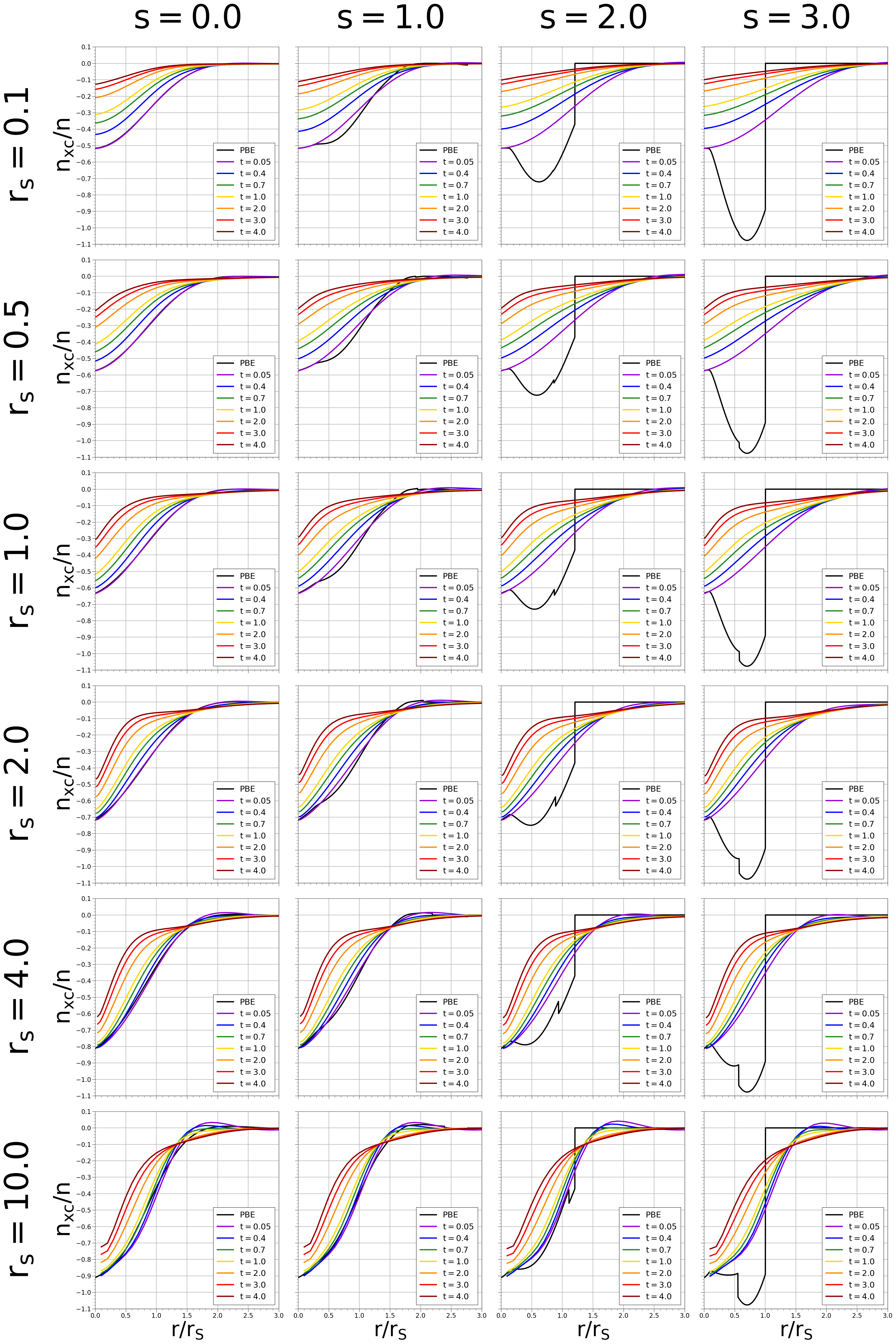}
    \caption{Grid of plots depicting temperature-dependent XC hole densities for unpolarized systems of various $r\s$, $s$, and $t$. We use black to denote the ground-state PBE XC hole density, while the CP-DFT XC hole at different reduced temperatures is depicted in various colors.}
    \label{fig:XCHoles.z_0.All}
\end{figure*}

\begin{figure*}
    \centering
    \includegraphics[width=0.84\linewidth]{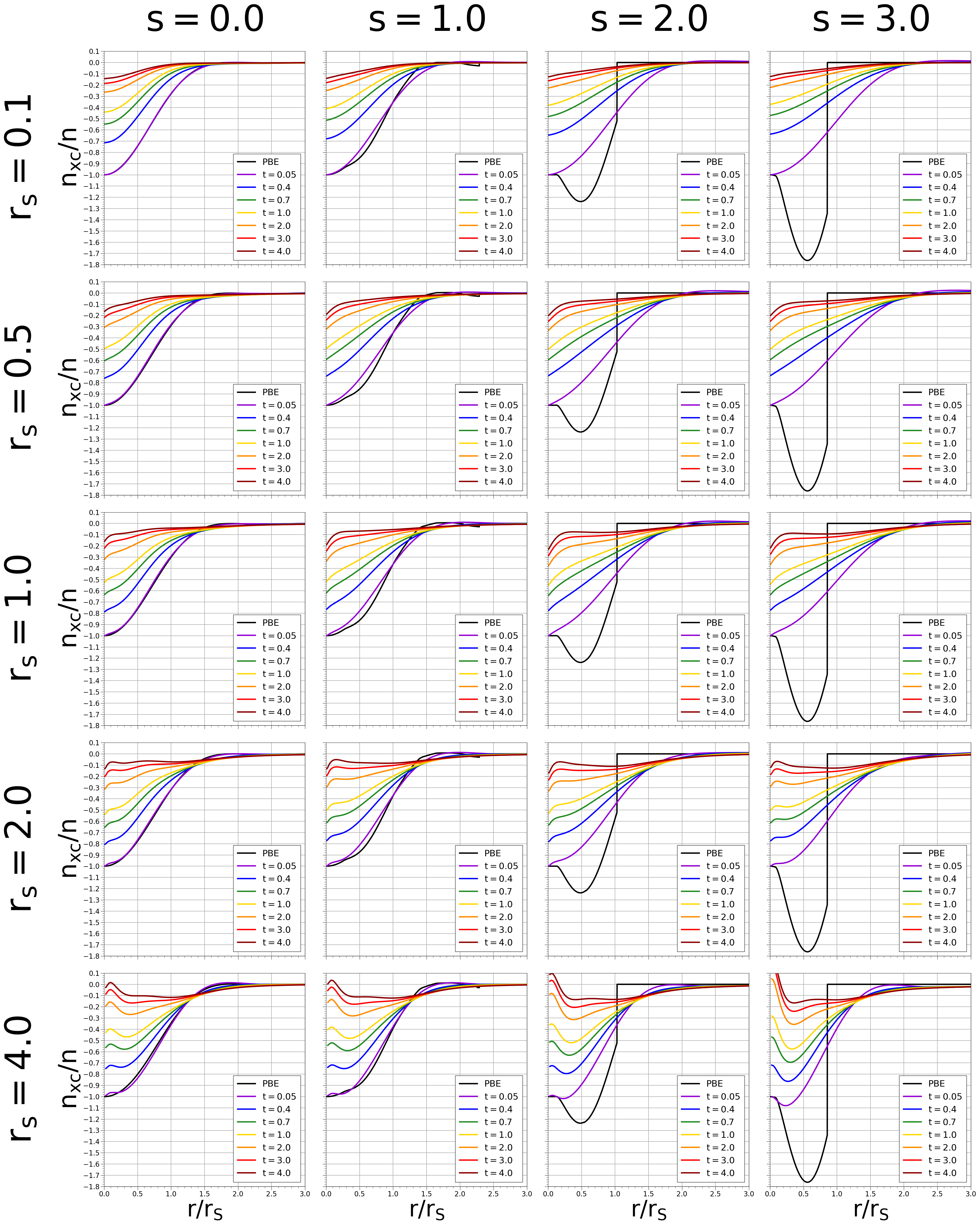}
    \caption{Grid of plots depicting temperature-dependent XC hole densities for fully polarized systems of various $r\s$, $s$, and $t$. We use black to denote the ground-state PBE XC hole density, while the CP-DFT XC hole at different reduced temperatures is depicted in various colors.}
    \label{fig:XCHoles.z_1.All}
\end{figure*}

Below we depict various plots illustrating the temperature-dependent XC hole densities output from the CP-DFT procedure outlined in the main text. Fig.~\ref{fig:XCHoles.z_0.All} depicts the results for unpolarized systems, while Fig.~\ref{fig:XCHoles.z_1.All} depicts the results for fully polarized systems. In all cases, the CP-DFT procedure produces an XC hole density with the same on-top value and energy as PBE as $t \rightarrow 0$.

\clearpage
\sec{Exact Conditions at Nonzero Temperature}

\ssec{Coordinate Scaling Inequality}

\begin{figure}[h]
    \centering
    \includegraphics[width=\linewidth]{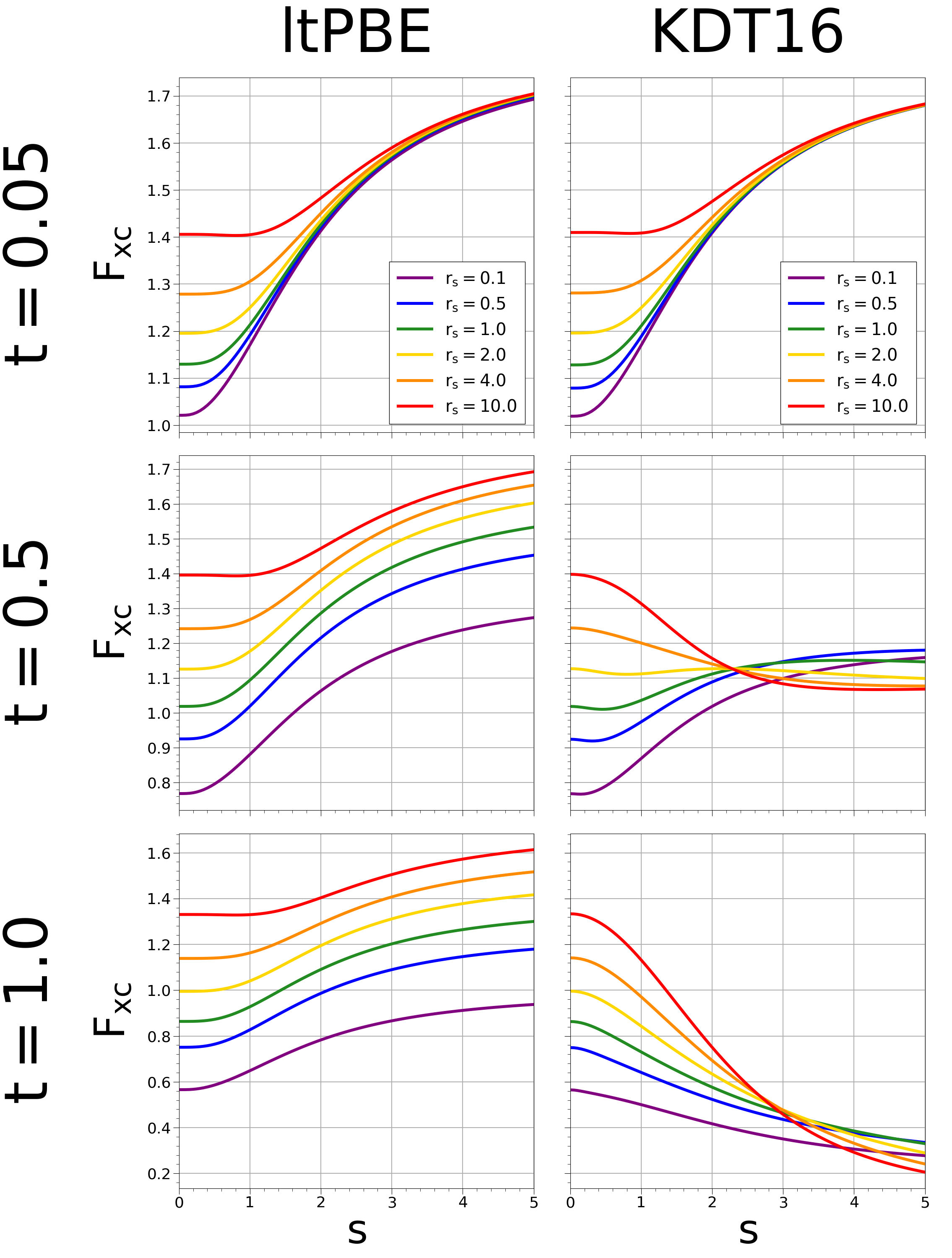}
    \caption{Temperature-dependent XC enhancement factors plotted as functions of the dimensionless gradient $s$ for unpolarized systems of various $r\s$ values, with each row having a fixed reduced temperature $t$. The left column depicts the ltPBE approximation, while the right column depicts KDT16. The thermal coordinate scaling inequality mandates these curves not cross for any $t$.}
    \label{fig:CSIgrid}
\end{figure}

Here we provide additional plots depicting the satisfaction of the thermal coordinate scaling inequality (Eq.~8 of the main text) by ltPBE, and violation by KDT16. In Fig.~\ref{fig:CSIgrid} we plot $F\xc$ curves at fixed values of $t$ for ltPBE and KDT16, illustrating the differences between the two approximations. The ltPBE enhancement factor, having inherited the curvature of the ground-state $F\xc\PBE$, naturally satisfies this exact condition at all temperatures. In contrast the KDT16 approximation only satisfies this exact condition when $t\rightarrow0$, where it recreates PBE, but violates it as the temperature is increased into the warm dense matter regime.

\ssec{Concavity Condition}

\begin{figure}[h]
    \centering
    \includegraphics[width=\linewidth]{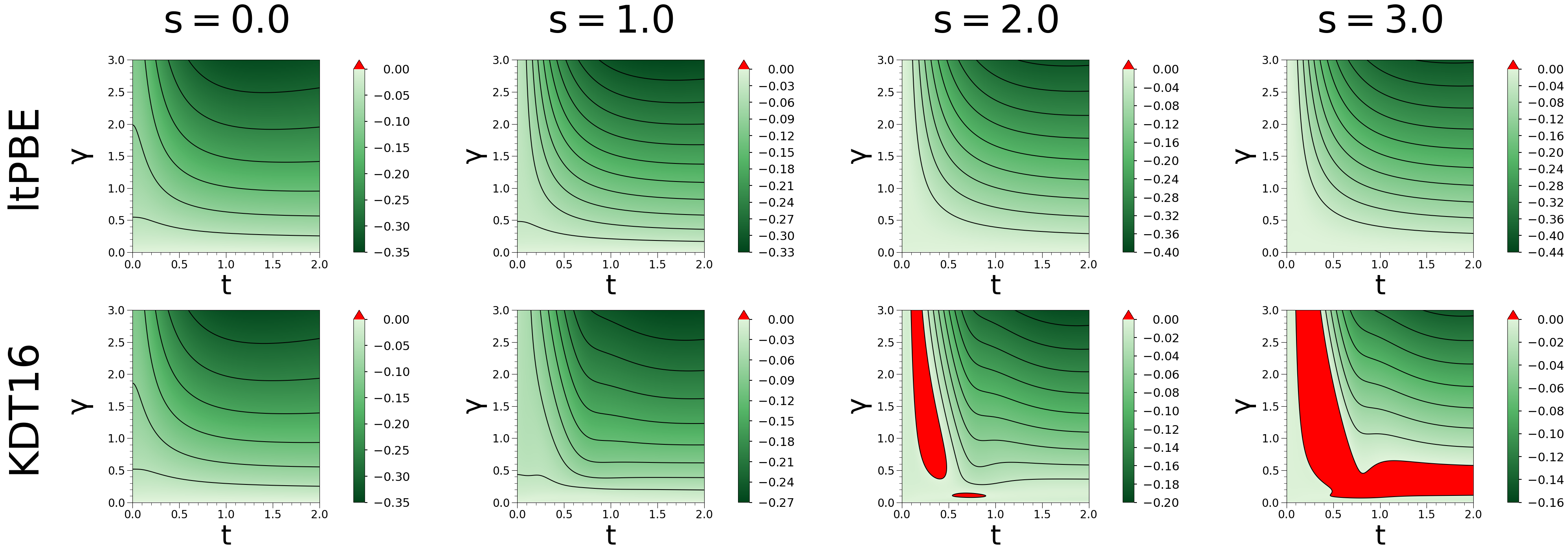}
    \caption{Contours of the thermal concavity condition inequality (Eq.~11) with initial density $n=1$. Here we show both the ltPBE (top) and KDT16 (bottom) approximations, confirming that the condition is satisfied by ltPBE, but is violated by KDT16 (red denotes positive).}
    \label{fig:CCfull}
\end{figure}

Here we show various contour plots depicting the thermal concavity condition (Eq.~11 of the main text) for ltPBE and KDT16 at different $s$ values. Looking at Fig.~\ref{fig:CCfull}, it is clear that the ltPBE approximation satisfies this condition, regardless of the value of $s$, due to its expression always being negative. It is also clear that KDT16 satisfies the concavity condition as $s \rightarrow 0$ (where it reduces to thermal LDA), but violates the condition as $s$ is increased. Noting that the density of a system scales according to $n_\gamma (\br) = \gamma^3 n (\gamma \br)$, we observe violations for KDT16 for both moderate/high densities with nonzero $s$ in the warm dense matter regime.

